\documentclass[12pt]{article}
\usepackage{amssymb,amsfonts,amsmath}
\begin{document}
 \hfill\break
\begin{center}\vspace{2.0cm}{\bf Hamilton-Jacobi Quantization of Landau-Ginzburg Theory}\\

\vspace{1.0cm}{\bf Walaa. I. Eshraim}\\

New York University Abu Dhabi, Saadiyat Island, P.O. Box 129188, Abu Dhabi, U.A.E\\
wie2003@nyu.edu
 $$$$
 $$$$
\end {center}
\begin{center}{\textit{keywords}: Hamilton-Jacobi
formalism, Singular Lagrangian, Path integral quantization of constrained systems.}\\
\end{center}
\begin{center}
\vspace{1.3cm}{\bf
 Abstract}\end{center} \vspace{0.4cm} We discuss the Hamilton-Jacobi approach for a constrained system. We obtain the equation of motion for a
 singular system as total differential equations in many variables. We investigate 
 the integrability conditions without using any gauge fixing condition. 
 The path integral quantization for systems with finite degrees of freedom is
applied to the field theories with constraints. So, we apply the Hamilton-Jacobi quantization to obtain the path integral of 
 the Landau-Ginzburg theory.
 \vfill
\section{Introduction}
$\\$\indent The quantization of constrained Hamiltonian systems
can be achieved by means of operator methods \cite{1,2} or by path
integral
quantization \cite{3,4,5}.\\
\indent The quantization of a massive spin one particle became the
center of interest of physicists especially after the pioneering
work of Faddeev \cite{3} who introduced the path integral quantization
of singular theories which possess first-class constraints in
canonical gauge. Since first class constraints are generators of
gauge transformations, this will lead to the gauge freedom. In
other words, the equations of motion are still degenerate and
depend on the functional arbitrariness, one has to impose external
gauge constraints for each first-class constraint.\\
\indent Using Dirac's method, one can see that the gauge fixing is not always an easy task \cite{Eshraim1, Eshraim0a, Eshraim0}. But, if the canonical method or Hamilton-Jacobi approach \cite{7,8} is used, the gauge fixing is not necessary to analyze the singular systems \cite{6}. The Hamilton-Jacobi treatment of constrained
systems leads us to obtain the equations of motion as total differential equations in
many variables, as seen in Refs. \cite{Eshraim3, Eshraim3a}. By using Hamilton-Jacobi quantization method, the Path integral quantization has been obtained for several constraint systems: (i) the Scalar field coupled minimally to the vector potential \cite{Eshraim4}, (ii) the electromagnetic field coupled to a spinor \cite{Eshraim5}, (iii) the relativistic local free field theory \cite{Eshraim3a}, (iv) the scalar field coupled to two flavours Fermionic through Yukawa couplings \cite{Eshraim3a}.\\
\indent The aim of this paper is to treat the Landau-Ginzburg
theory that gives an effective description of phenomenon precisely
coincides with scalar quantum electrodynamics as constrained
system. The path integral quantization is obtained by using canonical
method.

\section{Hamilton-Jacobi Formalism Of Constrained Systems}
$\\$\indent In this section, we study the constrained systems by
using the canonical method \cite{7,8} and demonstrate the fact that the
gauge fixing problem is solved naturally. The starting point of
this method is to consider the lagrangian $L\equiv
L(q_{i},\dot{q_{i}},\tau),\> i=1,2,\ldots,n$ with Hessian matrix
\begin{equation}\label{1}
A_{ij}=\frac{\partial^{2}L(q_{i},{\dot{q}}_{i},
\tau)}{{\partial{\dot{q}}_{i}}\>{\partial{\dot{q}}_{j}}}, \qquad
{i,j = 1,2,\ldots,n},
\end{equation}
of rank $(n-r)$, $r<n$. Then $r$ momenta are dependent. The
generalized momenta $p_{i}$ corresponding to the generalized
coordinates $q_{i}$ are defined as
\begin{align}
p_{a}&= \frac{\partial{L}}{\partial{\dot{q}}_{a}},\qquad {a =
1,2,\ldots,n-r}, \label{2}\\
p_{\mu}&= \frac{\partial{L}}{\partial{\dot{q}}_{\mu}},\qquad {\mu
= n-r+1,\ldots,n}.\label{3}
\end{align}
enable us to solve Eq. (2) for ${\dot{q}}_{a}$ as
\begin{equation}\label{4}
{\dot{q}}_{a}={\dot{q}}_{a}(q_{i},p_{a},{\dot{q}}_{\mu};\tau)\equiv\omega_a.
\end{equation}
Substituting Eq. (4), into Eq. (3), we get
\begin{equation}\label{5}
p_{\mu}=
\frac{\partial{L}}{\partial{\dot{q}}_{\mu}}\bigg|_{{\dot{q}}_{a}
=\omega_a}\equiv-H_{\mu} (q_{i},{\dot{q}}_{\mu},p_{a};t).
\end{equation}
Relations (5) indicate the fact that the generalized momenta
$p_{\mu}$ are not independent of $p_{a}$ which is a natural result
of the singular nature of the lagrangian.\\
 The canonical Hamiltonian $H_{0}$ is defined as
\begin{equation}\label{6}
H_{0}=-L(q_{i},{\dot{q}}_{\mu},{\dot{q}}_{a}\equiv\omega_a;\tau)+
p_{a}{\dot{q}}_{a}+ p_{\mu}{\dot{q}}_{\mu}\big|_{p_\mu=-H_\mu}.
\end{equation}
The set of Hamilton-Jacobi Partial Differential Equations is
expressed as
\begin{equation}\label{7}
H'_{\alpha}\bigg(\tau,~q_{\mu},~q_{a},~P_{i}=\frac{\partial
S}{\partial q_{i}} ,~P_{0}=\frac{\partial
S}{\partial\tau}\bigg)=0,\>\alpha=0,n-p+1,\ldots,n,
\end{equation}
where
\begin{equation}\label{8}
H'_{\alpha}= p_{\alpha}+H_{\alpha}~.
\end{equation}
The equations of motion are obtained as total differential
equation in many variables as follows:
\begin{align}
dq_{r}&=\,\frac{\partial H'_{\alpha}}{\partial p_{r}}\,
dt_{\alpha},\quad \quad\,\,\, r= 0,1,\ldots,n \, ,\label{9}\\
dp_{a}&=-\frac{\partial H'_{\alpha}}{\partial q_{a}}\,
dt_{\alpha},\;\qquad a=1,\ldots,n-p,\label{10}\\
dp_{\mu}&=-\frac{\partial H'_{\alpha}}{\partial
q_{\mu}}\,dt_{\alpha}, \,\;\quad \quad \alpha =
0,n-p+1,\ldots,n,\label{11}
\end{align}
\begin{equation}\label{12}
dz= \bigg(-H_{\alpha}+p_{a}\frac{\partial H'_{\alpha}}{\partial
p_{a}}\bigg) dt_{\alpha}.
\end{equation}
where $Z=S(t_{\alpha},q_{a})$. The set of equations (9-12) are
integrable if
\begin{equation}\label{13}
dH'_{\alpha}=0,\,\;\quad \quad \alpha=0,n-p+1,\ldots,n.
\end{equation}
If conditions (13) are not satisfied identically, one may consider
them as new constraint and a gain test the integrability
conditions, then repeating this procedure, a set of conditions may
be obtained.\\
In this case of integrable system, the path integral
representation may be written as \cite{9,10,11,12}.
\begin{equation}\label{14}
\left<Out\mid S\mid In\right>  =\int\prod_{a=1}^{n-r} dq^{a}
dp^{a} \exp{\left[ i\int_{t_{\alpha}}^{t'_{\alpha}}{\left(
-H_{\alpha}+ p_{a}\frac{\partial H'_{\alpha}}{\partial p_{a}}
\right)} dt_{\alpha}\right]}.
\end{equation}
One should notice that the integral (14) is an integration over
the canonical phase space coordinates $q_{a},p_{a}$.

\section{The Landau-Ginzburg theory}
\indent The Landau-Ginzburg theory that gives an effective
description of phenomenon precisely coincides with scalar quantum
electrodynamics is described by the lagrangian
\begin{equation}\label{15}
\mathcal{L}=-\frac{1}{4}\>F_{\mu\nu} F^{\mu\nu}+
(D_{\mu}\varphi)^\ast D^{\mu}\varphi-k \varphi^{\ast}
\varphi-\frac{1}{4}\lambda (\varphi^{\ast} \varphi)^2,
\end{equation}
where
\begin{equation}\label{16}
D_{\mu}\varphi=\partial_{\mu}\varphi-ieA_{\mu}\varphi.
\end{equation}
In the landau-Ginzburg theory $\varphi$ describes the cooper
pairs. In usual quantum electrodynamics, we would put $k=m^2$,
where $m$ is the effective mass of electron.\\
The Lagrangian function (15) is singular, since the rank of the
Hessian matrix
\begin{equation}\label{17}
A_{ij}=\dfrac{\partial^{2}L}{\partial{\dot{q}}_{i}
\partial{\dot{q}}_{j}}.
\end{equation}is three.\\
The canonical momenta are defined as
\begin{equation}\label{18}
\pi^{i}=\frac{\partial L}{\partial {\dot{A}}_{i}}=-F^{0i},
\end{equation}
\begin{equation}\label{19}
\pi^{0}= \frac{\partial L}{\partial{\dot{A}}_{0}}=0,
\end{equation}
\begin{equation}\label{20}
p_{\varphi}= \frac{\partial L}{\partial
\dot{\varphi}}=(D_{0}\varphi)^{\ast}=\dot{\varphi}^{\ast}+ieA_{0}\varphi^{\ast},
\end{equation}
\begin{equation}\label{21}
p_{\varphi^{\ast}}= \frac{\partial L}{\partial
{\dot{\varphi}}^{\ast}}=(D_{0}\varphi)=\dot{\varphi}-i\,e\,A_{0}\,\varphi,
\end{equation}
\indent From Eqs. (18), (20) and (21), the velocities
${\dot{A}}_{i},{\dot{\varphi}}^{\ast}$ and $\dot{\varphi}$ can be
expressed in terms of momenta $\pi_{i}, p_{\varphi}$ and
$p_{\varphi^{\ast}}$ respectively as
\begin{equation}\label{22}
{\dot{A}}_{i}=-\pi_{i}-\partial_{i}A_{0},
\end{equation}
\begin{equation}\label{23}
{\dot{\varphi}}^{\ast}=p_{\varphi}-ieA_{0}{\varphi}^{\ast},
\end{equation}

\begin{equation}\label{24}
{\dot{\varphi}}=p_{{\varphi}^{\ast}}+ieA_{0}{\varphi}.
\end{equation}
The canonical Hamiltonian $H_{0}$ is obtained as
\begin{multline}\label{25}
\qquad
H_{0}=\frac{1}{4}\>F^{ij}F_{ij}-\frac{1}{2}\>\pi_{i}\pi^{i}+\pi^{i}\,\partial_{i}A_{0}+p_{{\varphi}^{\ast}}p_{\varphi}+ieA_{0}{\varphi}p_{\varphi}
\\\,\,\qquad\qquad-ieA_{0}{\varphi}^{\ast}p_{{\varphi}^{\ast}}-(D_{i}\varphi)^{\ast}(D^{i}\varphi)
+k{\varphi}^{\ast}\varphi+\frac{1}{4}\lambda({\varphi}^{\ast}\varphi)^2.
\end{multline}\\
Making use of (7) and (8), we find for the set of HJPDE
\begin{equation}\label{26}
H'_{0}=\pi_{4}+H_{0},
\end{equation}
\begin{equation}\label{27}
H'=\pi_{0}+H=\pi_{0}=0,
\end{equation}
Therefor, the total differential equations for the characteristic
(9-11) obtained as
\begin{multline}
\quad \quad \quad \quad\qquad\qquad\quad dA^{i}=\frac{\partial
H'_{0}}{\partial\pi_{i}}\>dt+\frac{\partial
H'}{\partial\pi_{i}}\>dA^0,\\
=-(\pi^{i}+\partial_{i}A_{0})\,dt,\quad \quad \quad \quad \quad
\quad \quad \quad \quad \,\,\,\,\,\label{28}
\end{multline}

\begin{equation}\label{29}
dA^{0}=\frac{\partial H'_{0}}{\partial\pi_{0}}\>dt+\frac{\partial
H'}{\partial\pi_{0}}\>dA^0=dA^0,
\end{equation}
\begin{multline}
\quad \quad \quad \quad\qquad\qquad\quad d\varphi=\frac{\partial
H'_{0}}{\partial p_{\varphi}}\>dt+\frac{\partial
H'}{\partial p_{\varphi}}\>dA^0,\\
=(p_{{\varphi}^{\ast}}+ieA_{0}\varphi)\,dt,\quad \quad \quad \quad
\quad \quad \quad \quad \quad \,\,\,\,\,\label{30}
\end{multline}
\begin{multline}
\quad \quad \quad \quad\qquad\qquad\quad
d\varphi^{\ast}=\frac{\partial H'_{0}}{\partial
p_{\varphi^{\ast}}}\>dt+\frac{\partial
H'}{\partial p_{\varphi^{\ast}}}\>dA^0,\\
=(p_{\varphi}-ieA_{0}\varphi^{\ast})\,dt,\quad \quad \quad \quad
\quad \quad \quad \quad \quad \,\,\label{31}
\end{multline}
\begin{multline}
\quad \quad \quad \quad d\pi^{i}=-\frac{\partial H'_{0}}{\partial
A_{i}}\>dt-\frac{\partial H'}{\partial A_{i}}\>dA^0,\\
=[\partial_{l}F^{li}+ie(\varphi^{\ast}\partial^{i}\varphi+\varphi\,\partial_{i}\varphi^{\ast})+2e^{2}A^{i}\varphi\varphi^{\ast}]\,dt,\quad
\quad \quad \,\,\label{32}
\end{multline}
\begin{multline}
\quad \quad \quad \quad \quad \quad\,\,\,\,
d\pi^{0}=-\frac{\partial H'_{0}}{\partial
A_{0}}\>dt-\frac{\partial H'}{\partial A_{0}}\>dA^0,\\
=[\partial_{i}\pi^{i}+ie\varphi^{\ast}p_{{\varphi}^{\ast}}-ie\varphi\,p_{\varphi}]\,dt,\quad
\quad \quad \quad \quad \quad \quad\, \label{33}
\end{multline}
\begin{multline}
\quad \quad \quad \quad \quad \quad\,\,\,\,
dp_{\varphi}=-\frac{\partial H'_{0}}{\partial
\varphi}\>dt-\frac{\partial H'}{\partial \varphi}\>dA^0,\\
=[(\overrightarrow{D}\cdot\overrightarrow{D}\varphi)^{\ast}-k\varphi^{\ast}-\frac{1}{2}\lambda\varphi{\varphi^{\ast}}^2-ieA_{0}p_{\varphi}]\,dt,
 \,\,\,\,\,\label{34}
\end{multline}
and
\begin{multline}
\quad \quad \quad \quad \quad \quad\,\,\,\,
dp_{\varphi^{\ast}}=-\frac{\partial H'_{0}}{\partial
\varphi^{\ast}}\>dt-\frac{\partial H'}{\partial \varphi^{\ast}}\>dA^0,\\
=[(\overrightarrow{D}\cdot\overrightarrow{D}\varphi)-k\varphi-\frac{1}{2}\lambda\varphi^{\ast}{\varphi^2}+ieA_{0}p_{\varphi^{\ast}}]\,dt.
\quad \label{35}
\end{multline}
\indent The integrability condition $(dH'_{\alpha}=0)$ implies
that the variation of the constraint $H'$ should be identically
zero, that is
\begin{equation}\label{36}
dH'=d\pi_{0}=0,
\end{equation}
which leads to a new constraint
\begin{equation}\label{37}
H''=\partial_{i}\pi^{i}+ie\varphi^{\ast}p_{{\varphi}^{\ast}}-ie\varphi\,p_{\varphi}=0.
\end{equation}
Taking the total differential of $H''$, we have
\begin{equation}\label{38}
dH''=\partial_{i}d\pi^{i}+iep_{{\varphi}^{\ast}}d\varphi^{\ast}+ie\varphi^{\ast}dp_{{\varphi}^{\ast}}-ie\varphi\,dp_{\varphi}-iep_{\varphi}\,d\varphi=0.
\end{equation}
Then the set of equation (28-35) is integrable. Therefore, the
canonical phase space coordinates $(\varphi,p_{\varphi})$ and
$(\varphi^{\ast},p_{\varphi^{\ast}})$ are obtained in terms of
parameters $(t,A^{0})$.\\
\indent Making use of Eq.(12) and (25-27), we obtain the canonical
action integral as
\begin{multline}\label{39}
\quad Z=\int
d^{4}x(-\frac{1}{4}\>F^{ij}F_{ij}-\frac{1}{2}\>\pi_{i}\pi^{i}+p_{\varphi}p_{\varphi^{\ast}}+\overrightarrow{D}\varphi^{\ast}\cdot\overrightarrow{D}\varphi+k|\varphi|^2+\frac{1}{4}\lambda (\varphi^{\ast} \varphi)^2),
\end{multline}
where
\begin{equation}\label{40}
\overrightarrow{D}=\overrightarrow{\bigtriangledown}+ie\overrightarrow{A}.
\end{equation}
Now the path integral representation (14) is given by
\begin{multline}\label{41}
\left<out|S|In\right> = \int\prod_{i}\,dA^{i}\,d\pi^{i}\,
d\varphi\,
dp_{\varphi}\,d\varphi^{\ast}\,dp_{\varphi^{\ast}}\>exp\,\bigg[i\bigg\{\int d^{4}x\\
\bigg(-\frac{1}{2}\>\pi_{i}\pi^{i}-\frac{1}{4}\>F^{ij}F_{ij}+p_{\varphi}p_{\varphi^{\ast}}
+(D_{i}\varphi)^{\ast}(D_{i}\varphi)-k\varphi^{\ast}\varphi-\frac{1}{4}\lambda
(\varphi^{\ast} \varphi)^2\bigg)\bigg\}\bigg].
\end{multline}
\section {Conclusion}
\indent In this paper, the Landau-Ginzburg theory gives an
effective description of phenomenon precisely coincides with
scalar quantum electrodynamics. The Landau-Ginzburg has been quantized by constructing
a path integral quantization within the canonical method to
constrained system. The equations of motion are obtained as total
differential equations in many variables. All the constraints
coming from the Hamiltonian procedure and the integrability
conditions have been derived. The path integral quantization is
performed by using the action which is given by canonical method. The
integration is taken over the canonical phase space.


\begin{thebibliography}{99}
\bibitem{1} P.A.M. Dirac, lectures of Quantum Mechanics, Yeshiva University Press, New york (1964).
\bibitem{2} P.A.M. Dirac, Can J. Math. 2, 129 (1950).
\bibitem{3} L.D. Faddeev, Teoret. Mat. Fiz. {\bf 1}, 3 (1969)[Theor. Math. Phys. {\bf 1}, 1 (1970)].
\bibitem{4} L.D. Faddeev  and V. M. Popov, phys. Lett. {\bf B24}, 29 (1967).
\bibitem{5} S. I. Muslih and Y. G\"{u}ler, Nuovo Cimento {\bf B112}, 97 (1997).
\bibitem{Eshraim1}
W.~I.~Eshraim and N.~I.~Farahat,
%``Hamilton-Jackobi approach to the relativistic local free field with linear velocity of dimension D,''
Hadronic J. \textbf{29}, no.5, 553 (2006).
\bibitem{Eshraim0a}
W.~I.~Eshraim and N.~I.~Farahat,
%``Hamilton-Jacobi formulation of the scalar field coupled to two flavours Fermionic through Yukawa couplings,''
Islamic University Journal \textbf{15}, no.2, 141-151 (2007).
\bibitem{Eshraim0}
W.~I.~Eshraim and N.~I.~Farahat,
%``Hamilton-Jacobi treatment of Lagrangian with fermionic and scalar field,''
Rom. J. Phys. \textbf{53}, 437-444 (2008).
\bibitem{7} Y. G\"{u}ler, Nuovo Cimento  {\bf B107}, 1389 (1992).
\bibitem{8} Y. G\"{u}ler, Nuovo Cimento  {\bf B107}, 1143 (1992).
\bibitem{6} S. I. Muslih and Y. G\"{u}ler, Nuovo Cimento {\bf B113}, 277 (1998).
\bibitem{Eshraim3}
W.~I.~Eshraim and N.~I.~Farahat,
%``Hamilton-Jacobi formulation of a non-Abelian Yang-Mills theories,''
Electron. J. Theor. Phys. \textbf{5}, no.17, 65-72 (2008).
\bibitem{Eshraim3a}
W.~I.~Eshraim,
%``Hamilton-jacobi Treatment of Superstring and Quantization of Fields With Constraints,''
Alg. Groups Geom. \textbf{35}, no.4, 365-388 (2018).
\bibitem{Eshraim4}
W.~I.~Eshraim and N.~I.~Farahat,
%``Quantization of the Scalar Field Coupled Minimally to the Vector Potential,''
Electron. J. Theor. Phys. \textbf{4}, no.14, 61-68 (2007).
\bibitem{Eshraim5}
W.~I.~Eshraim and N.~I.~Farahat,
%``Path integral quantization of the electromagnetic field coupled to a spinor,''
Electron. J. Theor. Phys. \textbf{6}, no.22, 189-196 (2009).
\bibitem{9} S. I. Muslih, Nuovo Cimento {\bf B115}, 1 (2000).
\bibitem{10} S. I. Muslih, Nuovo Cimento {\bf B115}, 7 (2000).
\bibitem{11} S. I. Muslih, Nuovo Cimento {\bf B117}, 4 (2002).
\bibitem{12} S. I. Muslih, Mod. Phys. Lett. A {\bf A19}, 151 (2004).



\end{thebibliography}
\end{document}